\pdfoutput=1 
\documentclass{amcs}

\usepackage{soul}

\title[A Three-Level Parallelisation Scheme]{A Three-Level Parallelisation Scheme and\\ Application to the Nelder-Mead algorithm}

\author[ad1][ad2]{Rima~KRIAUZIEN{\. E}}
\author[ad1][]{Andrej~BUGAJEV}
\author[ad1][]{Raimondas~{\v C}IEGIS}

\correspondingauthor{Rima~KRIAUZIEN{\. E}}

\address[ad1]{
Department
of Mathematical modelling\\
Vilnius Gediminas Technical University,
Lithuania,
Sauletekio ave. 11, 
LT-10223 Vilnius\\
 e-mails: \url{rima.kriauziene@vgtu.lt}, \url{andrej.bugajev@vgtu.lt}, \url{raimondas.ciegis@vgtu.lt}
}
\address[ad2]{
Institute of Data Science and Digital Technologies\\
Vilnius university, Akademijos str. 4, LT-08663 Vilnius.
}

\Runauthors{R. Kriauzien{\. e}, A. Bugajev and Raimondas {\v C}iegis}

\newtheorem{rem}{Remark}

\begin{document}
\begin{abstract}
	We consider a three-level parallelisation scheme. The second and third levels define a classical two-level parallelisation scheme and some load balancing algorithm is used to distribute tasks among processes. It is well-known that for many applications the efficiency of parallel algorithms of the second and third level starts to drop down after some critical parallelisation degree is reached. This weakness of the two-level template is addressed by introduction of one additional parallelisation level. As an alternative to the basic solver some new or modified algorithms are considered on this level. The idea of the proposed methodology is to increase the parallelisation degree by using less efficient algorithms in comparison with the basic solver. As an example we investigate two modified Nelder-Mead methods. For the selected application, 
	a few partial differential equations are solved numerically 
	on the second level, 
	and on the third level the parallel Wang's algorithm is used 
	to solve systems of linear equations with tridiagonal matrices.
	A greedy workload balancing heuristic is proposed, which is
	oriented to the case of a large number of available processors. The complexity estimates of the computational tasks are 
	model-based, i.e. they use empirical computational data.
\end{abstract}

\maketitle

\section{Introduction}\label{sec:introduction}
Current trends in supercomputing show that in order to accumulate high
computing power,
computers with more, but not faster, processors are used. This trend
induces changes in the development of parallel algorithms. The important challenge
is to develop parallelization techniques which enable exploitation of substantially
more computational resources than the standard existing methods.

This paper deals with problems that can be split into
a collection of independent subproblems and this splitting step is repeated iteratively. The solutions of subproblems define the solution of an initial problem.
Thus, an additional splitting step increases the potential
parallelisation degree of a parallel algorithm.

Any multi-level parallelisation can be considered as a way to generate 
a pool of tasks. After the pool of tasks is obtained, it is not important 
how many parallelisation levels were used. 
However, often such final simplification of the template 
leads to a loss of an important information and as a consequence to degraded efficiency 
of the parallel algorithm.
Especially this is true if different levels of the scheme are 
characterised by different properties of an algorithm that should be properly 
addressed.

\begin{figure}[ht]
\includegraphics[width=0.48\textwidth]{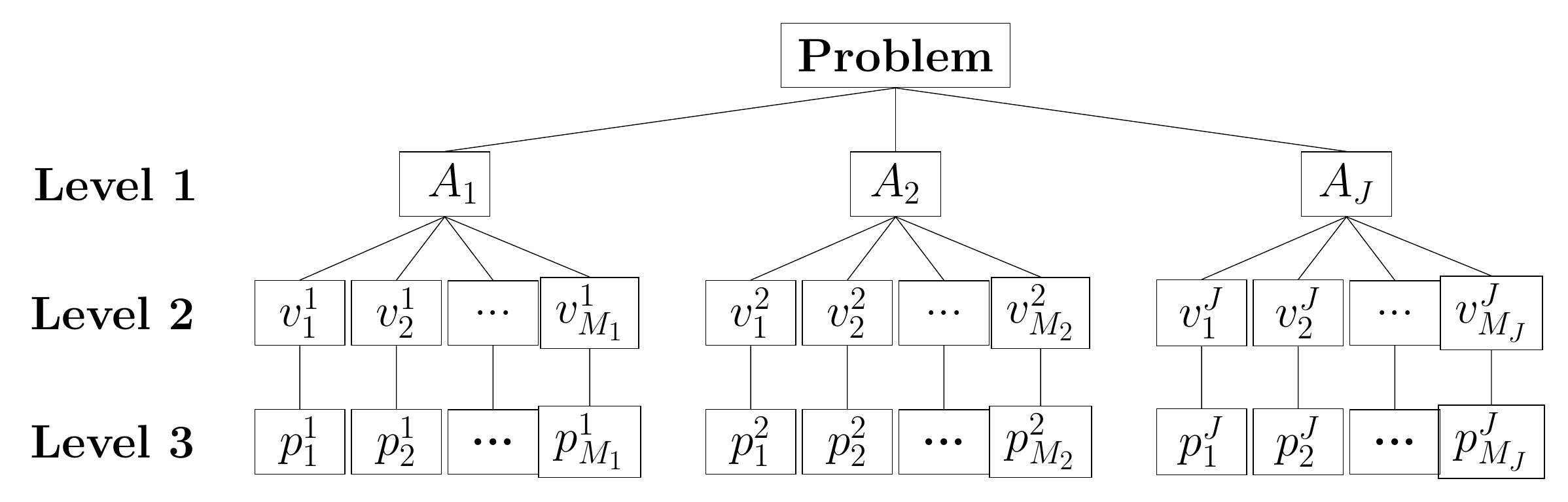}
\caption{Three level parallelisation scheme.}\label{pp}
\end{figure}

In this paper, we consider a special case of a three level parallelisation. The 
template of this approach is given in Fig.~\ref{pp}:
\begin{itemize}
\item
At the first level of parallelisation we assume that there exist a few parallel 
alternatives $A_j$ (see Figure~\ref{pp}) to the original modelling algorithm. The first level of parallelisation becomes a part 
of a new parallel algorithm and the degree of the first level parallelism can be selected 
dynamically during the computations -- a selection of the best algorithm is performed. In this paper as an example we consider two new
parallel modifications of the Nelder-Mead method \cite{nelder1965simplex}. 
\item
On the second level, a set of computational tasks $V^j = \{v_1^j, v_2^j , \ldots, v_{M_j}^j \}$ (see Figure~\ref{pp}) with different 
computational complexities is defined. These tasks are solved in parallel.
As an example we investigate the case
when computation of one value of the objective function requires to solve 
numerically  $M$ partial differential 
equations. The computational complexities of tasks are non-equal because different discretisation steps 
must be used for different equations in order to achieve the same accuracy
for each equation.
\item
The third level defines parallel algorithms to solve tasks from 
the second level. As an example we use Wang's algorithm to parallelise 
the solution of systems of linear equations with tridiagonal matrices
\cite{wang1981parallel}.
\end{itemize}

The second and the third levels define a well-investigated 
two-level parallelisation template. We note that load balancing techniques for
two-level parallelisation are  widely used in applications, 
see, e.g., \cite{Baravykaite}, \cite{Huismann}. 

The scheduling problem can be formulated representing a parallel 
algorithm by a directed acyclic graph (DAG). The vertices define computational 
tasks, the edges define connections/order among tasks. Then a set of partially 
ordered computational tasks is scheduled onto a multiprocessors system to 
minimise the computational time (or to optimise some other performance
criteria). It is well known that the scheduling problem is NP complete. 
Many interesting heuristics are proposed to solve it, we  
mention greedy algorithms \cite{silko_LNCS2002}, genetic algorithms 
\cite{sharma2015}, \cite{Singh}, simulated annealing 
and tabu search algorithms \cite{Kirkpatrick671}, \cite{tabu1}, \cite{tabu2}. Such 
algorithms include a possibility of dynamic scheduling and allow for 
tasks to arrive 
continuously and they can consider variable in time computational
resources.   

A scheduling task can be very challenging due to specificity of a given application
problem and the necessity to parallelise it on modern parallel 
architectures. As an example we mention the particle simulation which is 
solved by appropriate domain decomposition techniques \cite{Furuichi}. Another example is the  dynamic load balancing on heterogeneous clusters 
for parallel ant colony optimisation \cite{Llanes}. In the recent work \cite{datta2019exploiting} it is concentrated on the problem of high-dimensionality of the data while solving subspace clustering problem.

In this article we focus on the scheduling problem, when all tasks 
in the set are independent and can be solved in parallel.
It is well known that the given optimisation
problem can be redefined as a problem  to equalise the computational 
times of all processes. The simplest
load balancing algorithm is based on the assumption that the 
computation time is proportional to sizes of sub-tasks. Then the 
domain decomposition algorithm is  applied to guarantee that
the sizes of subtasks 
scheduled for each group of processors are equal \cite{Baravykaite}.

The quasi-optimal distributions of tasks   
can be obtained using the greedy strategy
to distribute the work on demand, i.e. to apply dynamic load 
balancing techniques such as work-stealing \cite{Imam}, 
self-organising process rescheduling \cite{Righi}.


However, the efficiency of two-level approach is
limited due to a typical saturation of the speed up of parallel
algorithms for increased numbers of processors and fixed sizes of tasks.
Exactly this situation has motivated us to introduce 
an additional  level
of parallelisation template. In most cases the usage of it 
leads to a less efficient algorithms than the initial state-of-the-art
algorithm. But the additional degree of parallelism on the second level
gives a large overall speed-up, if the number of available resources is large.

Recent developments of new architectures
of parallel processors make even more challenging
the task 
to build accurate theoretical  performance models. The empirical data 
shows that for some advanced algorithms the efficiency of parallel 
computations can depend non-monotonically on the size of a task.
Thus the model-based
load balancing method starts to become the main tool in developing 
efficient and accurate task scheduling algorithms. In our work 
we build the model for prediction of computation time empirically by solving 
the specialised benchmarks for a wide range of problem sizes and numbers 
of processors. In fact this analysis resemblance the classical
experimental  strong scalability 
analysis of a given parallel algorithm. 
We note, that these measurements are always done for all processes working
simultaneously in order to reflect their actual performance during the 
execution of real applications (see, also \cite{Lastovetsky2,Lastovetsky1}).

Here we mention two interesting papers, where the model-based task
 scheduling algorithms are considered. 
In \cite{Lastovetsky1}, it is concentrated on multicore co-processors Xeon Phi, where the empirical computation time curves are
used to find optimal parameters for a workload distribution. 
The obtained model predicts non-monotonic dependence of computation 
speed on the sizes of problems. 
The authors call their approach "load imbalancing", however, it can be
considered as an advanced balancing which adapts the scheduling
algorithm to the specificity of Xeon Phi processors. Obviously in this case 
the assumption that computation time is proportional  to the  
task size is not valid.
In a similar research \cite{Lastovetsky2}, computations
were performed on non-uniform memory access (NUMA) parallel platform
with various shared on-chip resources such as Last Level Cache.
Again the model-based approach enables to take 
into account the specific properties of the algorithm and processors.   
The  matrix multiplication and Fast Fourier Transform are used
as benchmark problems. 
It is interesting to note that, according to the presented results,
the globally optimal solutions may not load-balance the sizes of sub-tasks. 
The authors pay a special attention to the energy efficiency of 
calculations. We note, that there are some papers that are specifically 
dedicated to load balancing of energy efficiency \cite{Perez}. In our work 
we formulate some restrictions that are connected to energy efficiency as 
well -- we do not use additional available computational resources if the 
parallelisation efficiency drops below some specific level.
The other work \cite{Lastovetsky3} is dedicated to model-based optimisation on hybrid heterogeneous systems 
composed of CPUs and accelerators. In that research authors investigate the problem of communications costs due to uneven workload 
distribution between accelerators and CPUs. They propose to generalise the $\tau$-Lop \cite{tauLop} model for heterogeneous computations. 

In this paper we propose a general methodology for parallelisation of
algorithms. As an example we use it to solve some applied
optimisations problems. 
is shown
The superiority of the three level parallelisation scheme
is shown, 
comparing it with two level paralleisation scheme. On the second level a set of different-size tasks is defined, which
is a typical situation for computation of one value of a black box  objective 
function. In most cases these tasks (or groups of tasks) are independent 
but computationally costly. Thus each task also should
be solved in parallel.
This fact leads to a necessity of the third level.  
The second and third levels of the template define a set of tasks solved in parallel and
some load balancing algorithm should be used to take into account the different sizes
of subtasks. 
The necessity of the additional  first level comes from the assumption of
having more computational resources than can be utilised by the two-level parallelisation
approach. It is a consequence  
of the efficiency saturation for parallel algorithms when the size of the problem
is fixed and the number of processes is increased. 
We select a different optimization method (or a modification of the basic solver)
which gives additional degrees of parallelisation thus enabling
the possibility to use more processors. 
At the first level of the template the optimal algorithm
is selected. This part  requires to find a 
compromise between the increased parallelisation degree and the decreased convergence 
rate of the modified parallel optimization algorithm.


{
In this work we are also interested to address some green computing (GC) challenges.
In a broader sense GC is the practices and procedures of designing, manufacturing, using of computing
resources in an environment friendly way while maintaining overall computing performance and finally disposing in a way that
reduces their environmental impact \cite{Saha2018}. 
The research in green computing is done in many areas \cite{nemalikanti2011achieving}:
Energy Consumption; E-Waste Recycling; Data Center Consolidation and Optimization; Virtualization; I. T Products and Eco-labeling.
One of approaches for optimisation of energy consumption on the software level is the autotuning software, which is able to optimise its own execution parameters with respect to a
specific objective function (usually, it is execution time) \cite{carretero2015energy}. Well known examples of autotuning software are: FFTW \cite{frigo2005design} (fast Fourier transformations); ATLAS \cite{whaley1998automatically}, PHiPAC \cite{bilmes1997optimizing} (dense matrix computations); OSKI \cite{vuduc2005oski}, SPARSITY \cite{im2001optimizing} (sparse matrix computations).
}

{
 Usually, the goal for any autotuning software is to achieve the same 
result with the same resources, however, reducing the computation
time -- in terms of parallelisation it means to increase the 
parallelisation efficiency. Another way to decrease the power consumption 
is to increase the efficiency by avoiding inefficient calculations; 
this may slightly increase the execution time, however will give 
a reasonable increase of parallel 
efficiency, which leads to the energy savings. 
We propose to control the efficiency of the parallel algorithm
 on the load balancing stage of the parallelization template. In many 
cases this strategy reduces  the amount of computational resources used 
in  computations. This analysis is done a priori, meaning that the user 
knows how many cores should be used for solving
a specific parallel task even before starting real computations.
}

This paper makes the following contributions:
\begin{enumerate}
\item
We propose to extend the typical two level parallelisation, which is
usually accompanied by some load balancing technique, by adding one additional level. Also, we investigate  the possibility
to limit the number of used processors 
to sustain the parallelisation efficiency at the selected level. This approach  let us to avoid the inefficient calculations, supporting the green computing technology.

{As an example two different families of parallel Nelder-Mead methods were investigated: 
the family of the generalised parallel Nelder-Mead method \cite{lee2007parallel} and the parallel versions of the classical Nelder-Mead method. 
In order
to perform the load balancing on the second and third levels of the 
proposed template,
we use the complexity model of tasks which is based on the computational data (also known as model-based), 
as it is done in recent state-of-the-art 
works \cite{Lastovetsky1},\cite{Lastovetsky2},\cite{Lastovetsky3}.
We demonstrate a big potential of this new technique.
}

\item
A parallel version of the Nelder-Mead method is proposed, which does not change the convergence properties of the sequential optimisation algorithm. We note, that there were some attempts to parallelise 
this optimisation method before \cite{lee2007parallel}, \cite{Klein2014}. 
However, in these papers the convergence properties are changed and these
changes are not studied comprehensively enough. 
Moreover, it is questionable whether these parallel algorithms
 are applicable in the case of
 small-dimension problems. 

{Our parallel algorithm leads to an increasement of 
the parallelisation degree 
up to factor three. However, the introduced changes do not affect 
the convergence of the sequential optimisation algorithm. The experimental comparison of this new parallelization 
algorithm with the state-of-the-art technique \cite{lee2007parallel} 
is provided. The obtained experimental results show that in the case of
the Rosenbrock function the convergence properties of the parallel
 algorithm  \cite{lee2007parallel}  are much worse than of the classical
sequential Nelder-Mead algorithm.    
}

\end{enumerate}

The rest of this paper is organised as follows. In Section~\ref{sec:workload} 
the workload balancing problem is formulated, the selection of the optimal algorithm is provided and a general strategy for 
workload distribution is presented along with the efficient 
workload distribution algorithm. 
In Section~\ref{sec:pa}  
the detailed description of three parallelisation levels are given
for the studied case. We consider the approximation of boundary conditions of Schr{\" o}dinger equation.
The modified Nelder-Mead method is used
to solve local optimisation problems on the first level, on the 
second level a set of partial differential equations are solved
numerically, 
and on the third level Wang's algorithm is used to 
solve systems of linear equations in parallel. In Section~\ref{sec:ne} the results of 
computational experiments are provided and the efficiency of the
proposed three-level parallelisation template is analysed.  In Section~\ref{section6} the comparison of different Nelder-Mead parallelisation methods is presented.
The final conclusions are done in Section~\ref{sec:conclusions}.

\section{Workload balancing problem} \label{sec:workload}

In this section we formulate the workload balancing problem for 
the two level parallelisation. Also we present a greedy 
scheduling algorithm to distribute the processes among tasks. Next, we introduce the additional level -- the first and second levels of the two level parallelisation technique become the second and the third levels, accordingly and the first level is a new parallelisation level. On the first level the selection of the optimal algorithm is performed. 

{First, we will present two level parallelisation template. Assume that we solve a given problem by using the basic method $A$. The solution process consists of $K$ blocks of tasks (a simple DAG)}
\begin{equation}\label{Atasks}
  A = \{ V_1 \prec V_2 \prec \ldots \prec V_K \},
\end{equation}
and all blocks must be solved sequentially one after another. 
Each block consists of $M$ tasks
$$
   V_k = \{ v_1(X_k), v_2(X_k), \ldots,  v_M(X_k) \}, \quad k=1, \ldots, K, 
$$
where $X_k$ defines a set of parameters for the $V_k$ block. $V_k$ defines the first level of two level parallelisation scheme. 
Each task $v_m$ can be solved by parallel algorithm -- this is the second level of the scheme.

The complexities of tasks $v_m$ are different, however, they are known in advance and do not depend on $k$.
For each task $v_m$ the prediction of computation
time $t_m(p)$, $p \leq P$,
$m=1, \ldots, M$
is given -- it is based on the modelling results, $P$ is the number of processors in a  parallel system. We assume that up to $P_m$ processes the computation 
time monotonically decreases: 
\begin{equation}\label{pcon0}
     t_m(p_2) < t_m(p_1), \quad \mbox{for} \;\;  p_1 < p_2 \leq P_m. 
\end{equation}
For $ P_m$ the predicted computation time function $t_m(p)$ reaches
the minimum value:
\begin{equation}\label{pcon1}
   t_m (p) \geq t_m({P}_m), \quad \;\; p >  {P}_m.
\end{equation}

\begin{figure}[ht]
\centering
\includegraphics[width=0.49\textwidth]{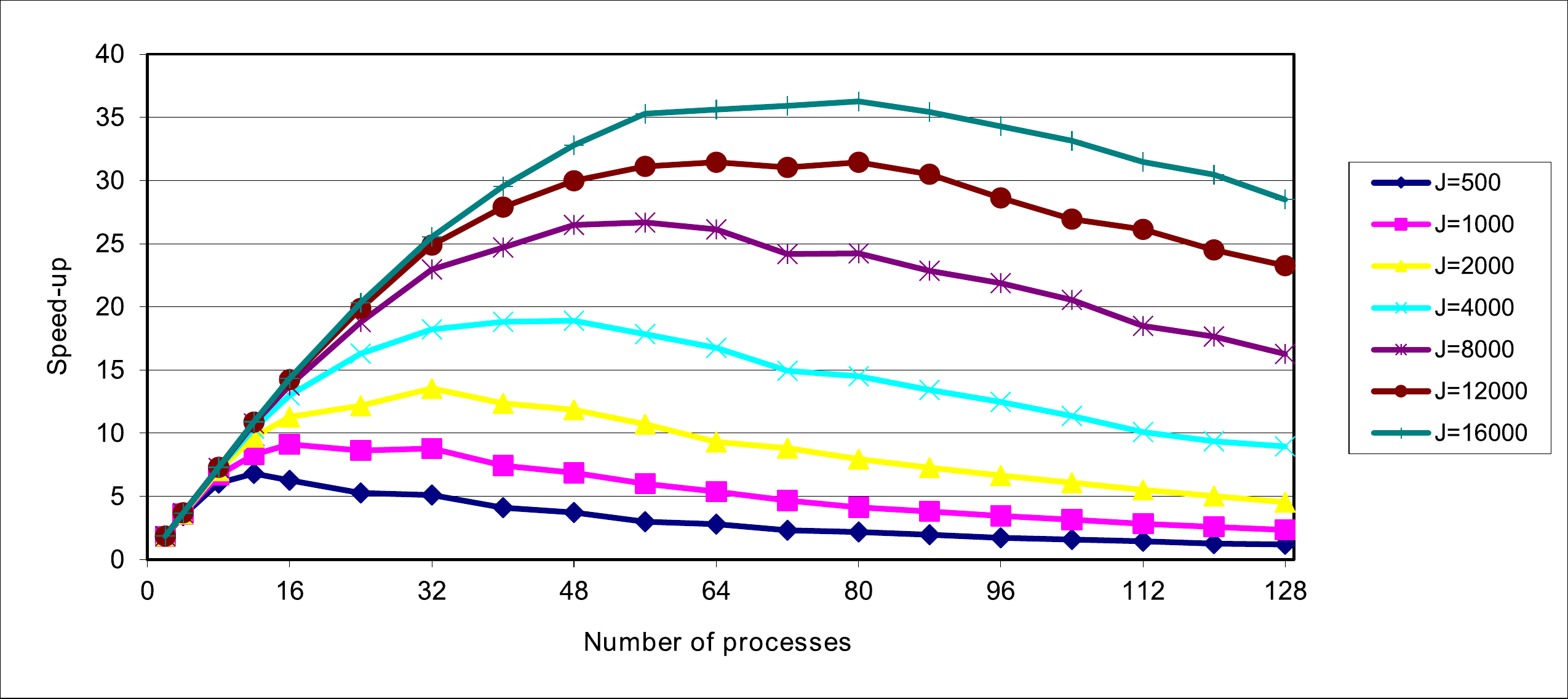}
\caption{The speed-ups of Wang's parallel algorithm for different 
number of processes $p$ and sizes $J$ of systems. The detailed specification of processors is presented in Section~\ref{sec:pa}.}\label{Speedup}
\end{figure}

Such a model of computation time $t_m(p)$ is important for algorithms with limited scalability such as Wang's
algorithm. In  Fig.~\ref{Speedup} we present speed-ups of this 
algorithm for different sizes of linear systems. It is important to
mention that the provided results include 
some additional costs for computation of the objective function  
along with Wang's algorithm 
computational costs. These additional calculations slightly 
increase the overall parallelisation scalability, 
thus the provided figure represents the optimistic scenario for general Wang's
algorithm and the realistic scenario for actual computations, that were done in this paper.

In our specific case this data was derived from a simple benchmark implementing Wang's algorithm. This benchmark performs computations using different numbers of processes and different problem complexity parameters $J$. It is important to note, that nodes were artificially loaded with calculations to imitate the real situation. For example, with the number of processes $p=4$ there were 32 tasks that were solved by 128 processes at the same time. Thus this benchmark must be run once, using all processes available.

From Figure~\ref{Speedup} it follows that the computation time monotonically decrease till some 
critical number of processes and therefore the efficient usage of 
processes is limited to this number of processes.   
Even for large size systems, when the  number of equations is
 $J = 16000$, the 
maximum number of processes $P_m$ does not exceed 80. 
This analysis justifies our motivation 
to use the multi-level approach in order to solve the given applied problem. 

In the two-level parallelisation scheme  for each block of tasks $V_k$ we select the number of processes such that
the overall solution time is minimised:
$$
   \arg \min_ {(p_1, \ldots, p_M) \in Q} \max_{1 \leq m \leq M} t_m (p_m),  
$$
where a set of feasible processors distributions $Q$ is defined as
$$
  Q = \{ (p_1, \ldots, p_M):  p_1 + \ldots + p_M \leq P \}.
$$

\begin{rem}
In the case when we solve only few large size tasks and the remaining tasks
are much smaller and the number of processes $P$ is not very big, 
the optimal scheduling is obtained when a few 
smaller tasks are combined  into one group $\widetilde v_m$. Then sub-task $\widetilde v_m$ 
consists of tasks $v_{l_1}, \ldots, v_{l_n}$. The computation time 
for this combined task is predicted by the model:
$$
   \widetilde t_m(p_m) = \sum_{i=1}^n t_{l_i}(\widetilde p_{l_i}), \quad
    \widetilde p_{l_i} = \min \left( p_m, P_{l_i} \right). 
$$  
In this work we are interested to solve the scheduling problem, when the number
of processes is large, so the aggregation step is not used.
\end{rem}

Next, we propose a simple greedy partitioning algorithm, 
which is described in Algorithm~\ref{balal}.
It aims to find a near-optimal distribution
of $M$ tasks of different sizes  between homogeneous $P$
processes by using the model-based complexity model $t_m(p)$ (similar 
ideas are also used in \cite{Lastovetsky2}).
We assume that $P \geq M$. 
The interesting feature of the presented algorithm 
is that for a given number of processes $P$
the number of active processes can be taken less than $P$
to minimise the overall execution time of the parallel algorithm.

The algorithm starts from the initial distribution when one process is
assigned
for each task and the predictions of parallel execution times 
are calculated using the selected performance model. 
Then, the greedy iterative procedure is applied to distribute
the remaining processes. 
At each iteration, one additional process is assigned to the   
task which has the largest predicted computation time. Then its 
parallel execution time is updated. 
Iterations are repeated until all processes are distributed or
the number of processes for some task reaches the limit $P_m$.

\begin{algorithm}[!h]
\caption{The algorithm  for distribution of $P$ processes between $M$ tasks}\label{balal}
\begin{algorithmic}[1]
	\STATE Set  $p[m]=1$, for $m=1, \ldots, M$
	\STATE $P=P-M$
	\STATE Compute $t_m(p[m])$, for $m=1, \ldots, M$
	\STATE stop = 0
	\WHILE{ $P>0 \; \& \,$ stop $==0$}
		\STATE find $j$ such that  $ t_j(p[j]) =  \max\limits_{1\leq m \leq M} t_m(p[m])$
		\IF{ $p[j] ==  P_j$}
			\STATE stop = 1
		\ELSE
			\STATE $p[j]=p[j]+1$
			\STATE $P=P-1$
		\ENDIF
	\ENDWHILE
\end{algorithmic}
\end{algorithm}

Note, that before $t_m(p)$ has reached the minimum, value starts to decrease slowly, thus the parallelisation efficiency drops. Therefore, it may be wise
to restrict the number of processes by taking into account the
efficieny value.

We define the maximum number of processes $\widetilde{P}_k$ 
for which the efficiency condition is still satisfied
\begin{equation}\label{pcon2}
   E_p (V_k) \geq E_{min}, \quad \mbox{for} \;\; p \leq  \widetilde{P}_m,
\end{equation}
where $E_p(V_k) = S_p(V_k) /p$ ir $S_p(V_k) = t_k (1)/t_k (p)$,  $E_{min} \in [0,1]$ is a given efficiency lower bound. 
Estimate (\ref{pcon2}) is used to modify the limit of the maximum number of processes (\ref{pcon1}) that can be used to solve the $j$-th task
\begin{equation}\label{pcon3}
P_m = \min{({P}_m,\widetilde{P}_m)} .
\end{equation}

Therefore, in the presented technique $P_m$ includes two restrictions:
\begin{itemize}
\item
The number of processes cannot exceed the number after which the speed-up drops down (see Fig.~\ref{Speedup}).
\item
The number of processes is limited by  efficiency requirement 
(\ref{pcon2}), which states: the number of processes 
per block of tasks $V_k$ is not allowed to be increased if 
the efficiency of the parallel algorithm on the third level reaches the 
critical value $E_{min}$.
\end{itemize}

In fact the second level of the two-level scheme can be used alone, however, it is limited due to Amdahl's law \cite{amdahl}, i.e. the efficiency begin to drop as the number of processes increases for a fixed size of problem. 
Two-level approach let us to solve this issue up to some point.  

Exactly this situation has motivated us to introduce 
an additional level
of parallelisation template.

In the new three-level parallelisation scheme, the second and third levels represent the two-level scheme part described before. Additionally, we add new first level of the template. We assume that there exist parallel alternative algorithms $A_j$: 
$$
   A_j = \{ V_1^j \prec V_2^j \prec \ldots \prec V_{K_j}^j \}, \quad j=1, \ldots, J. 
$$
Each block $V_k^j$ consists of $M_j$ independent tasks 
$$
   V_k^j = \{ v_1^j (X_k), v_2^j (X_k), \ldots, v_{M_j}^j(X_k) \}, \quad k=1, \ldots, K.
$$ 
The numbers of blocks of tasks $K_j$, the numbers of tasks per block $M_j$, the sizes of tasks $ \vert v_m^j \vert $ may be different for different $j$. 

Next, we select the optimal algorithm according to the number of resources available. We denote
$$
T_P(A_j) = T_P(V^j) K_j
$$
the total solution time for algorithm $A_j$. The block of tasks $V^j$ is solved by using the heuristic proposed above. Then the optimal algorithm is defined as
$$
    \arg \min_{1 \leq j \leq J} T_P(A_j).
$$
The usage of $ j>1$ may
lead to a less efficient algorithm than the initial basic algorithm. But the additional degree of parallelism gives a large overall speed-up.

%
%

\section{Application of the three-level parallelisation scheme}\label{sec:pa}


First, we briefly present the problem which is used to test our methodology. 
We solve an initial-boundary value Schr{\" o}dinger problem
formulated in a finite space domain \cite{ShrodABC}:
\begin{equation}\label{eq2}
\begin{cases}
i\dfrac{\partial u}{\partial t} + \dfrac{\partial^2 u}{\partial x^2}=0,
 \quad  x\in (A,B), \;\; t\in (0,T], \\
u(x,0)=u_0(x), \quad x \in [A,B],\\
L_l u(A)=0, \quad L_r u(B)=0, \quad t\in (0,T],
\end{cases}
\end{equation}
where operators $ L_l, L_r $ define the nonlocal/transparent 
boundary conditions.

Let $\omega_h$ and $\omega_\tau$ be discrete uniform grids
with space and time steps $h$, $\tau$:
\begin{align}\label{diskr}
& \omega_h = \big\{ x_j: \; x_0 = A, \, x_J = B, \, x_{k} = x_{k-1} + h,
 \, k= 1, \ldots, J \big\}, \\
& \omega_\tau = \{ t^n: \;\;  t^n = n \tau, \quad n=0, \ldots, N,
 \, N \tau = T \}.
\end{align}
Let $U^n_{j}$ be a numerical approximation of
the exact solution $u_{j}^n=u(x_j, t^n)$ at the grid
points $(x_j, t^n) \in  \omega_h \times \omega_{\tau}$.
For functions defined on the grid we introduce
the forward and backward difference quotients with respect
to $x$
\begin{align*}
 & \partial_x U^n_{j} = {(U^n_{j+1} - U^n_{j})}/{h}, \quad
  \partial_{\bar x} U^n_{j} = {(U^n_{j} - U^n_{j-1})}/{h}
\end{align*}
and similarly the backward difference quotient and the averaging
operator with respect to $t$
\begin{align*}
 & \partial_{\bar t} U^n_j = {(U^n_{j} - U^{n-1}_{j})}/{\tau},
  \quad
 U_j^{n-0.5} = 0.5\left( U^n_j + U^{n-1}_j\right).
\end{align*}
We approximate the differential equation (\ref{eq2})
by the Crank-Nicolson finite difference sche\-me \cite{radziunas}
{\begin{align}
    & i\partial_{\bar t} U^n_j +  \partial_{x} \partial_{\bar x} U_j^{n-0.5} = 0,
\quad x_j \in \omega_h, \;\; n> 0. \label{fds1}
\end{align}}
A very interesting approach to construct the approximate local  artificial boundary conditions is based on approximation of the transparent boundary condition
$$\partial_n u + e^{-i\frac{\pi}{4}}D_t^{1/2}u=0$$
by rational functions. The discrete boundary conditions can be written as:
{
\begin{align} \label{eq:RFBC}
& \partial_n u=
-e^{-i\frac{\pi}{4}}\bigg(\bigg(\sum\limits_{k=0}^{l}a_k\bigg) u{-}\sum
\limits_{k=1}^{l}a_kd_k\varphi_k\bigg), \; x=a,b,
\end{align}
}
where $\partial_n u$ is the normal derivative, $\varphi_k$ are solutions of 
the initial value problem for ODEs 
\cite{ShrodABC}:
\begin{align*}
\dfrac{d\varphi_k(x,t)}{dt}+d_k\varphi_k(x, t)=u\left(x,t\right), 
\; x=A, B,
 \; k=1, \ldots, l.
\end{align*}
Our aim is to find optimal values of parameters  
$\{a_0,a_1,\dots a_l, d_1,a_2,\dots d_l\}$, 
when the following minimisation problem is solved
\begin{equation} \label{eq:L2M}
\begin{split}
E_\infty^c  =& \min \limits_{\{a_k, d_k \}}  V_k,
	\quad V_k=\max\limits_{1\leq m \leq {\tilde M}}^{} v_m(X_k), \\
v_m=&\max \limits_{j \in [0,J_m],
\, n \in [0,N_m]} \big|u(x_{j,m}, t^n_m) - U_{j,m}^n \big| ,
\end{split}
\end{equation}

and $ \widetilde M$ specially selected  benchmark PDEs are solved. 

In all examples we use $l=3$, i.e., the dimensionality of the optimization problem
\eqref{eq:L2M} is equal to 7. Here discrete approximations of PDEs represent the tasks $v_m$ in \eqref{Atasks}. To solve $v_m$ we must find solutions of $N$ systems of linear equation with tridiagonal matrix \cite{ShrodABC}. According to our three-level parallelisation scheme, the calculations of a single point in minimisation problem \eqref{eq:L2M} define the block of tasks $V_k$.

The systems of linear equations with tridiagonal matrices are solved 
using Wang's algorithm. It is well known that if the size 
of a system is $J$ and $p$ processes are used then the computation 
time can be estimated as 
\begin{equation}\label{compW}
T_{Wp} = 17 \dfrac{J}{p} + 8p + T_{c1}(p),
\end{equation}
where $T_{c1}(p)$  defines  communication costs. The time to
compute a value 
of the objective function $f$ for the specified equation can be
estimated as
\begin{equation}\label{compOF}
T_{Op} = c_1\dfrac{J}{p} + T_{c2}(p).
\end{equation}
In this work 
instead of theoretical complexity models (\ref{compW}) and (\ref{compOF}) 
we use $t_m(p)$, $m=1, \ldots, M$, based on empirical computations for a selected 
set of benchmark problems. 
Such an approach takes into account all specific details
of the parallel algorithm and the computer system.

It is interesting to note that the complexity of computational task 
$v_m$ depends on both parameters: 
the number of linear equations $J_m$ of the system and the number of integration in time 
steps $N_m$. 
The computation time $T_{mp}$ 
is equal to $N_m t_m (p)$, but the scalability of the parallel algorithm 
depends on $J_m$ only, since the integration in time is done sequentially step by step.

Next, we present an example with $M=4$, where 
four different benchmark PDE problems (\ref{eq2}) with explicit solutions \cite{szeftel,Zlotnik} are defined as:
\begin{enumerate}
\item
\begin{equation}\label{exmpl1}
u(t,x){=}\dfrac{\exp{ \left( -i\pi/4 \right)}}{\sqrt{4t-i}}\exp\left(
 \dfrac{ix^2{-}6x{-}36t} {4t-i} \right),
\end{equation}
$ x \in [-5,5]$, $ t\in [0,0.8]$. The problem is approximated on the  
uniform grid $J \times N = 8000 \times 4000$.

\item
\begin{equation}\label{exmpl2}
\begin{split}
u(t,x)=\dfrac{1}{\root{+}\of{1+i {t}/{\alpha}}} 
\exp\left(ik(x-x^{(0)}-kt)\right.
\\
\left.
 -\dfrac{(x-x^{(0)} -2kt)^2} {4(\alpha+it)} \right),
\end{split}
\end{equation}
where $k=100, \alpha=1/120, x^{(0)}=0.8$ . 
$x \in [0,1.5]$, $t \in [0,0.04]$. We use the
uniform discretisation grid  $ J \times N = 12000 \times 4000$.

\item
The solution is defined by (\ref{exmpl1}), $x \in [-10,10]$, $ t\in [0,2]$. We use the
uniform discretisation grid $J \times N = 16000 \times 10000$.

\item
The solution if defined by (\ref{exmpl2}),
where $k=100, \alpha=1/120, x^{(0)}=0.8$. 
$x \in [0,2]$, $t \in [0,0.08]$. We use the
uniform discretisation grid $ J \times N = 16000 \times 8000$.
\end{enumerate}

Next, we consider the problem \eqref{eq:L2M} as a local optimisation problem, which can be solved using an iterative algorithm  with a given initial starting point.
As a local optimiser Nelder-Mead algorithm is used \cite{nelder1965simplex}.


We propose a family of modifications of the original Nelder-Mead algorithm  in order to increase
the parallelisation degree of it. 

At each iteration the following four different scenarios can be obtained:
\begin{itemize}
\item Reflection -- compute the value $f_R$ of the objective function
   at the point $X_R$. Depending on the value $f_R$ this can be the end of the iteration.
\item Expansion -- depending on the $f_R$, an additional 
computation of the objective function
   at the point $X_E$ is done, meaning the total computation of two objective function values: $f_R, f_E$.
\item Contraction -- depending on the $f_R$, an additional 
computation of the objective function at point $X_C$ is done, meaning the total
computation of two objective function values: $f_R, f_C$.
\item Compression -- compute $m$ objective function values, as well as $f_R$ and $f_C$.
 Here $m$ is the number of simplex dimensions.
\end{itemize}

The first three scenarios require to compute one or two values of the objective function from the set: 
$f_R$, $f_E$, $f_C$. We can neglect the last scenario, because it occurs 
very rarely. For the first three scenarios we propose to compute two or 
three points simultaneously. Algorithmically this means that we change 
the order of computations, which let us to parallelise the Nelder-Mead method. 
In most cases only two of three points will be used. Therefore, 
some redundant calculations will be performed, however, this modification gives an 
additional parallelisation of computations. 

Thus, two modifications of the sequential ($A_1$) Nelder-Mead method are defined.
For $A_2$ we compute in parallel values $f_R, f_E$ and for $A_3=3$ we 
compute in parallel all three values $f_R, f_E, f_C$. 
As a test case we assume that the first scenario is relatively rare, 
the extension step is done with probability 2/3 and
contraction steps occurs with probability  1/3.
Then we get that the algorithmic efficiency of the proposed parallel modifications
are equal to $\gamma_2 = 0.75$ and $\gamma_3 = 2/3$, respectively. We note, that these 
values can be estimated more precisely for specific applications, and one example is 
given for the computational experiments with the Rosenbrock objective function in Section~\ref{section6}.

On the first level different parallel algorithms can be used, however, the proposed approach is oriented to the cases when the increased degree of parallelisation gives the speed-up at the cost of efficiency which is a typical situation in parallel algorithms theory (Amdahl's law). 
As one more example we mention  new algorithms  developed  to solve the global optimisation problems. The modification of the well-known DIRECT method \cite{finkel2003direct} was presented in \cite{Stripinis2018}, it is called DIRECT-GL. 
 The new modification is based on the idea at each iteration to analyse more potential optimal rectangles. This approach increases the global sensitivity of the method  but in many cases this property is achieved at the cost of additional computations. The potential parallelisation degree of DIRECT-GL algorithm can increase up to 2-3 times. But the results of computational experiments in \cite{Stripinis2018} show that for many benchmark problems (in \cite{Stripinis2018} these cases are numbered 1,2,5,6,20,21,22,24,35,37,38,47,48,49,52) the DIRECT-GL algorithm increases the computational costs to achieve the same accuracy of approximations as DIRECT algorithm.
 Thus, the classical DIRECT algorithm and its modification DIRECT-GL fit well into  the proposed three-level parallelisation template.
 Then the degree of parallelisation should be increased only if this increasement compensates the reduced efficiency of the modified algorithm.
 Thus we state, that in order to apply the proposed  three level parallelisation scheme, first the computations of one point should be parallelised by a two-level parallelisation approach. Then alternative cases of parallel algorithms with additional degrees of parallelisation should be identified and the optimal  algorithm should be selected.
 
%

\section{Experimental results} \label{sec:ne}

In this section we present results of the parallel scalability tests. 
All parallel numerical tests in this work were performed on the computer cluster
``HPC Saule\-te\-kis'' at the
High Performance Computing Center of Vilnius University, Faculty of Physics.
We have used up to 8 nodes with
Intel\textsuperscript{\textregistered} Xeon\textsuperscript{\textregistered}
processors E5-2670 with 16 cores (2.60 GHz) and 128 GB of RAM per node.
Computational nodes are interconnected via the InfiniBand network.

Our main goal is to investigate the efficiency of the proposed three
level template of
workload distribution between processes. First, we have selected three specific benchmarks with different discretizations (\ref{diskr}), when $M=4$ discrete approximations of PDEs
\eqref{fds1}
are solved numerically to compute one value of the objective function.
The sizes $(J_m \times N_m)$, $m=1, \ldots, 4$ 
of discrete problems are given in Table~\ref{benchmarks}.

\begin{table}[ht]
\caption{Benchmarks with different sizes $J_m \times N_m$ of the discrete
 problem \eqref{fds1}}\label{benchmarks}
\centering
\scalebox{0.9}{
\begin{tabular}{cccccc}
\hline
&Benchmark 1&&Benchmark 2&&Benchmark 3\\
\cline{2-2}\cline{4-4}\cline{6-6}\\[-0.3cm]
Eq. & Sizes && Sizes && Sizes \\
\hline	
1 & $8000 \times 40000$ && $8000 \times 20000$ && $8000 \times 10000$ \\
2 & $4000 \times 20000$ && $4000 \times 20000$ && $2000 \times 20000$ \\
3 & $2000 \times 20000$ && $4000 \times 10000$ && $2000 \times 10000$ \\
4 & $2000 \times 10000$ && $2000 \times 10000$ && $1000 \times 20000$ \\
\hline	
\end{tabular}}
\end{table}

In the first benchmark the size of one task $v_1$  is much bigger than 
sizes of the remaining three tasks.
In the second  benchmark two changes are done. They make this set
of tasks more suited for parallelisation on large number of processes:
the size of task $v_1$ is reduced twice by taking a smaller number of time
steps $N_1$; the size of task $v_3$ remains the same, but the number of 
points $J_3$ is increased twice, therefore the scalability of Wang's
algorithm is improved for this task. 
In the third benchmark the relative sizes of tasks $v_m$ are more homogeneous
than in the first benchmark, but this result is achieved by reducing the 
number of space grid points $J_2, J_4$, therefore the scalability of Wang's algorithm is decreased for these two tasks, especially for $v_4$.

First, we exclude the efficiency condition from the load balancing algorithm
by taking $E_{min}=0$ in (\ref{pcon2}).
The distribution of processors between tasks are presented in 
Tables~\ref{t:2}--\ref{t:4}. We also provide the actual computation 
time $T_p$ along with $T_{Mp}$ that were predicted by the theoretical
complexity model. As we can see from Table~\ref{t:2} the model and experimental times are close to each other. The experimental time is smaller in cases when there is no interpolation error. Also it is smaller than the model time -- it is expected result, model times (see Figure~\ref{Speedup}) are based on benchmark, that imitate pessimistic scenario -- as it was mentioned before, all nodes were artificially loaded at the same time. The prediction accuracy depends on many parameters such as cluster architecture, network loads during computations.

For comparison purposes we provide the results obtained by using the two-level parallelisation
template. $K=1$, then the first level of the three-level template is not used.

It is important to note, that in Tables~{\ref{t:2}-\ref{t:b1_1}} we present the CPU
time needed to compute one 
useful point (\ref{eq:L2M}), i.e., the actual time is divided by $\gamma_k\,k$, which represents the usefulness of computations. Optimal algorithm $A_k$ is selected automatically using the approach that was 
described above.

As it follows from Table~\ref{t:2}, the usage of the first level with $k=3$ and 
$P=128$ processes increases the potential speed-up from 38.75 to 60.44. If $P=128$
and $k=1$ then only 70 processes 
are used. However the result is very similar to the case when $P=64$ processes are used,
which means that these additional resources 
are used very inefficiently.

\begin{figure}[ht]
\centering
\includegraphics[width=0.23\textwidth]{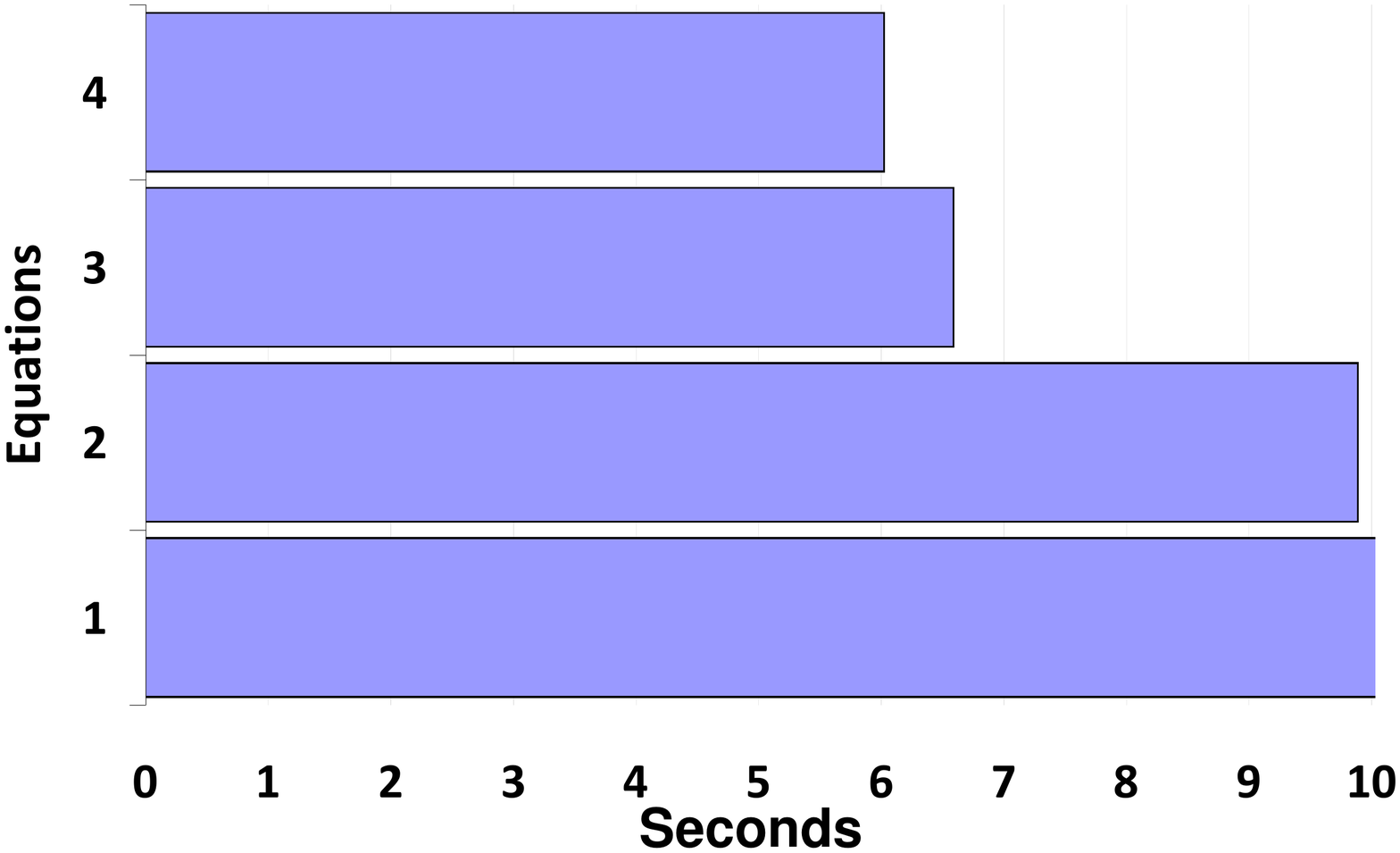}
\includegraphics[width=0.23\textwidth]{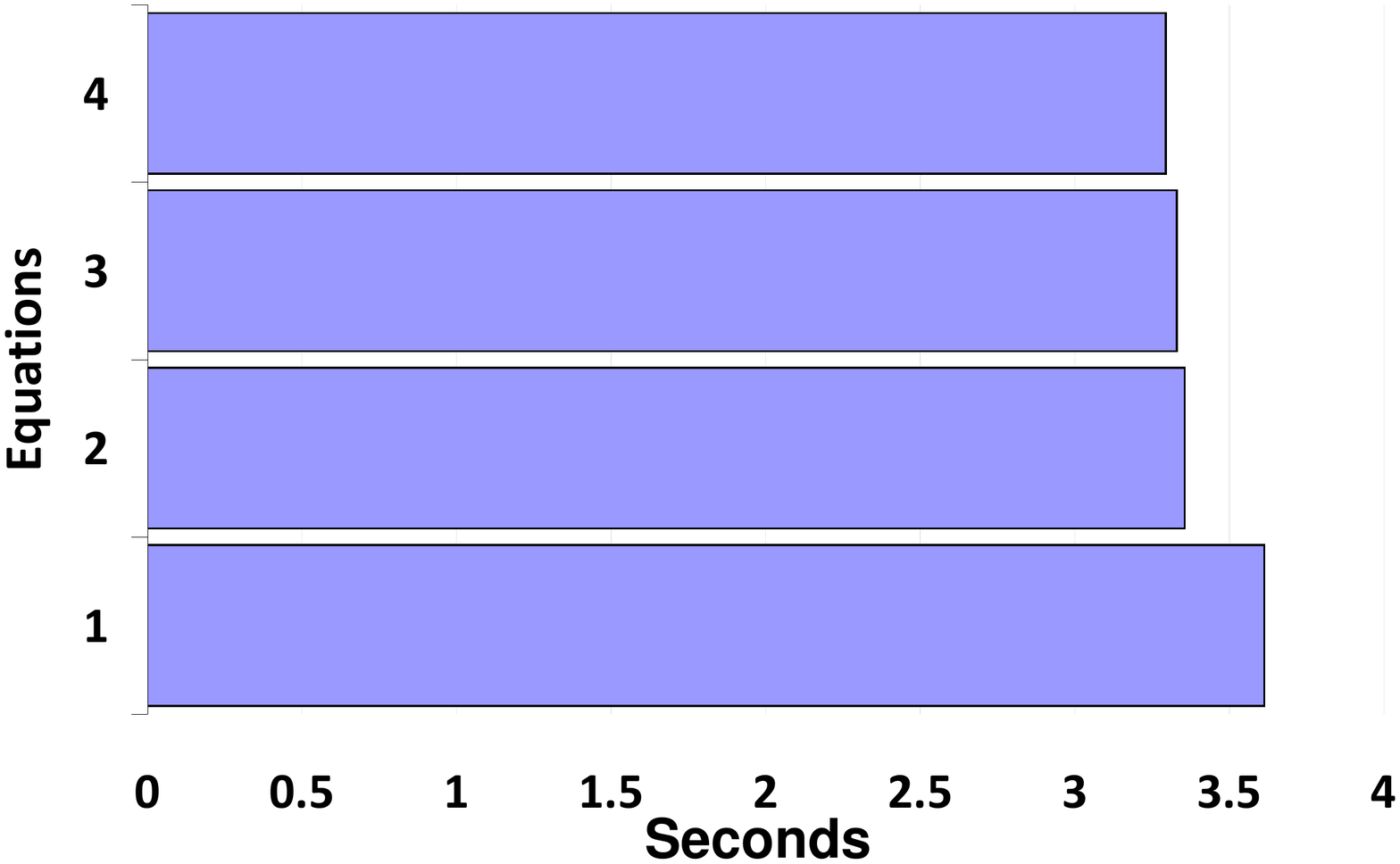}
\caption{Experimental model times for Benchmark 1 with $p=16$(left) and $p=64$(right)}\label{Gant1}
\end{figure}

In the Fig.~\ref{Gant1} the Gantt charts show theoretical model time $t_m(p)$, that is needed to obtain the 
solutions of different equations. The workload distribution becomes 
closer to uniform as the number of processes is increased.

\begin{table}[ht]
\caption{The results for Benchmark 1. $T_p$ is the CPU time in seconds required to compute one 
useful point (\ref{eq:L2M})}\label{t:2}
\centering
\begin{tabular}{rccc}
\hline
p& 16 & 32 & 64\\
\cline{2-4}\\[-0.3cm]
&\multicolumn{3}{c}{$k=1$}\\
\hline\\[-0.3cm]
Eq. 1	&	10	&	22	&	50\\
Eq. 2	&	3	&	5	&	8\\
Eq. 3	&	2	&	3	&	4\\
Eq. 4	&	1	&	2	&	2\\
\hline
Total number of $ p $ & 16	&	32	&	64\\

\hline
Model $T_{Mp}$	&	11.145	&	5.784	&	3.614	\\
$T_p$		&	11.003	&	5.394	&	3.608	\\
Speed-up		&	12.679	&	25.862	&	38.664	\\

\hline
\end{tabular}
\vskip 3mm
\begin{tabular}{rccccc}
\hline
p& 96 && 128&& 128\\
\cline{2-2}\cline{4-4}\cline{6-6}\\[-0.3cm]
&\multicolumn{1}{c}{$k=2$}&&\multicolumn{1}{c}{$k=3$}&&\multicolumn{1}{c}{$K=1$}\\
\hline\\[-0.3cm]
Eq. 1	&	34	&	&	29	&	&	56\\
Eq. 2	&	8	&	&	7	&	&	8\\
Eq. 3	&	4	&	&	4	&	&	4\\
Eq. 4	&	2	&	&	2	&	&	2\\
\hline
Total number of $ p $ & 96	& & 126	& &	70\\

\hline
Model $T_{Mp}$ 		&	2.742	&	&	2.272	&	&	3.605	\\
$T_p$		&	2.719	&	&	2.308	&	&	3.600	\\
Speed-up		&	51.307	&	&	60.444	&	&	38.75	\\
\hline
\end{tabular}
\end{table}

\begin{table}[ht]
\caption{The results for Benchmark 2. $T_p$ is the CPU time in seconds required to compute one 
useful point (\ref{eq:L2M}).}\label{t:3}

\centering
\scalebox{0.75}{
\begin{tabular}{rcccccccc}
\hline
p& 16 & 32 & 64 && 96 & 128&& 128\\
\cline{2-4}\cline{6-7}\cline{9-9}\\[-0.3cm]
&\multicolumn{3}{c}{$k=1$}&&\multicolumn{2}{c}{$k=2$}&&\multicolumn{1}{c}{$K=1$}\\
\hline\\[-0.3cm]
Eq. 1	&	9	&	18	&	37	&	&	26	&	37	&	&	56\\
Eq. 2	&	4	&	8	&	15	&	&	12	&	15	&	&	18\\
Eq. 3	&	2	&	4	&	8	&	&	7	&	8	&	&	8\\
Eq. 4	&	1	&	2	&	4	&	&	3	&	4	&	&	4\\
\hline
Total number of $ p $ & 16	& 32	&	64 &&96&128&&86\\

\hline
Model time		&	6.59	&	3.36	&	2.01	&	&	1.65	&	1.34	&	&	1.8	\\
$T_p$		&	6.69	&	3.37	&	1.98	&	&	1.62	&	1.33	&	&	1.86	\\
Speed-up		&	13.6	&	27.03	&	46.03	&	&	56.24	&	68.25	&	&	49.03	\\
\hline
\end{tabular}}
\end{table}

\begin{table}[!ht]
\caption{The results for Benchmark 3. $T_p$ is the CPU time in seconds required to compute one useful point (\ref{eq:L2M}).}\label{t:4}
\centering
\scalebox{0.75}{
\begin{tabular}{rcccccccc}
\hline
p& 16 & 32 & 64 && 96 & 128&& 128\\
\cline{2-4}\cline{6-7}\cline{9-9}\\[-0.3cm]
&\multicolumn{3}{c}{$k=1$}&&\multicolumn{2}{c}{$k=2$}&&\multicolumn{1}{c}{$K=1$}\\
\hline\\[-0.3cm]
Eq. 1	&	8	&	16	&	32	&	&	24	&	32	&	&	56\\
Eq. 2	&	4	&	8	&	16	&	&	12	&	16	&	&	31\\
Eq. 3	&	2	&	4	&	8	&	&	6	&	8	&	&	8\\
Eq. 4	&	2	&	4	&	8	&	&	6	&	8	&	&	9\\
\hline
Total number of $ p $ & 16	& 32	&	64 && 96 & 128 && 104\\
\hline
Model time		&	3.33	&	1.76	&	1.05	&	&	0.87	&	0.7	&	&	0.9	\\
$T_p$		&	3.38	&	1.76	&	1.06	&	&	0.86	&	0.7	&	&	0.95	\\
Speed-up		&	14.33	&	27.55	&	45.96	&	&	56.72	&	69.08	&	&	51.23	\\
\hline
\end{tabular}}
\end{table}

\subsection{The control of efficiency}
The reduction of the energy consumption is an important goal, especially when increasment of computation speed-up are small for additional processes. The presented results indicate 
that in some cases there is a highly inefficient usage of computational 
resources. 

For the purposes of controlling the efficiency of 
calculations the condition \eqref{pcon2} 
was introduced in Algorithm~\ref{balal}.  
This condition guarantees that the efficiency of the numerical solution of each 
block of tasks will be at least $E_{min}$. It is important to note, that we are not attempting to generate optimal 
mappings of processors -- we have developed an heuristic that 
provides the quality of distribution of tasks, that is sufficient for the most practical purposes.
The quality of the algorithm is improved when more processors are available.


Next, a more detailed analysis of the  Benchmark 1 is provided. 
In Table~\ref{t:b1_1} the results for $E_{min}>0$ are presented. 
Comparing the results in Table~\ref{t:b1_1} with 
the results 
in Table~\ref{t:2} we see that for 
$K=1$ and $E_{min}=0.6$ the number of processes for the first 
equation is decreased by $14$, however, the computation 
times are almost the same as it was in the case of $E_{min}=0$. Also, for $K=3$ 
the efficiency requirement begins to limit the number of 
processes for $E_{min} = 0.75$ and it decreases further with 
$E_{min} = 0.8$.

\noindent However, even then a three level approach with $K=3$ 
is superior to the standard two-level approach in terms of the final speed-up. The results in Table~\ref{t:b1_1} indicate that even 
for the efficiency limitation $E_{min}=0.75$ the proposed
three-level approach lets to maintain a big number of parallel processes
active, this number  is equal to $(26+7+4+2)\times 3 = 117$. The 
speed-up is 56 and the efficiency of the parallel algorithm  is
$56/117 \approx 0.48$. The last column in Table~\ref{t:b1_1} 
with $K=1$ presents the results for the two-level approach
(without the first level).
A straightforward two-level parallelisation 
approach would have the limited parallelisation possibility especially 
for problems of the size $J=2000$. For such small subproblems it would be 
possible to utilise only  up to 32 processes
(Fig.~\ref{Speedup}), the speed-up would be quite limited as well.

\begin{table}[!h]
\caption{The results for Benchmark 1 with $E_{min}>0$. $T_p$ is the CPU time in seconds required
 to compute one 
useful point (\ref{eq:L2M}).}\label{t:b1_1}
\centering
\begin{tabular}{rccccc}
\hline
p&  && 128\\
\cline{2-6}\\[-0.3cm]
$E_{min}$&\multicolumn{1}{c}{$0.75$}
&&\multicolumn{1}{c}{$0.8$}
&&\multicolumn{1}{c}{$0.6$}
\\ 
& \multicolumn{1}{c}{$k=3$}
&&\multicolumn{1}{c}{$k=3$}
&&\multicolumn{1}{c}{$K=1$}
\\ 
\hline\\[-0.3cm]
Eq. 1		&	26	&	&	19			&	&42	\\
Eq. 2		&	7	&	&	5			&	&8		\\
Eq. 3		&	4	&	&	3			&	&4		\\
Eq. 4		&	2	&	&	1			&	&2		\\
\hline
Total number of $ p $ & 117	&& 84	&& 56 \\
\hline
Model $T_{Mp}$	& 2.45	&	&	3.17	&	&3.84  \\
$T_p$	& 2.49	&	&	3.08	&	&3.76	 \\
Speed-up	& 56.03	&	&	45.37	&	&37.11	 \\
\hline
\end{tabular}
\end{table}

Note, that all previous results represents the analysis based on a single Nelder-Mead iteration. Next, we solve the actual real-world optimisation problem \eqref{eq:L2M}. The maximum number of processes $P=128$ 
the load balancing algorithm has selected $k=1$. The number of Nelder-Mead method iterations was fixed to 1000.
The parallel and sequentional versions gave the same results the minimum value of the error $E_{\infty}^C = 0.0806$. 
The sequentional version of computations took $180286$ seconds,
the parallel version computations took $2232$ seconds. Thus, a speed-up factor 
of $81.8$ was achieved. 
The selection of $k=1$ indicates that the number of processes  
can be greatly increased -- the algorithm has selected $k=1$ automatically for a
given number of processes. 

\section{The comparison of different Nelder-Mead parallelisation methods}\label{section6}
Here we present the analysis of the convergence properties of different modifications of the Nelder-Mead method. As it was mentioned before, the convergence rate of the selected
algorithm  directly affects the parallelisation efficiency, which 
is represented by $\gamma_k$, where $k$ is the parallelisation degree. 
In this section we measure $\gamma_k$ by measuring the experimental parallel efficiency of algorithms.

The detailed analysis of convergence behaviour for different objective functions is 
out of the scope of this research. However, the objective function from the previous sections is 
suitable for a narrow class of applications. Thus, to perform a comparison of different
parallel versions of Nelder-Mead method we minimise the Rosenbrock objective function that is widely used by 
researchers in the field of optimisation theory \cite{fajfar2018nelder}, \cite{Stripinis2018}. 

We show that in the case of the Rosenbrock function the real experimental $\gamma_k$ values 
are different than were assumed  to be in the experiments of the previous sections. 
The reason is that the significant number of iterations require to compute only one point 
$F_R$.

We compare the  results of our parallel modification of the Nelder-Mead method with  
 the state-of-art technique proposed in \cite{lee2007parallel}. 
As a benchmark we use the Rosenbrock function 
\begin{equation}\label{Ros}
f(x_1, \dots , x_d) = \sum\limits_{i=1}^{d-1} 100(x_{i+1}-x_i^2)^2+(1-x_i)^2,
\end{equation}
which makes the optimisation problem challenging. 
It should be noted that the parallel algorithm  \cite{lee2007parallel} can achieve 
the parallelisation degree $K$ that is equal to the optimisation problem dimension $d$.
Thus potentially this algorithm is well suited for parallel computers with a big number
of processes.

\begin{table}[ht]
\caption{The $\gamma_k$ values for direct Nelder-Mead parallelisation} \label{musu}
\centering
\begin{tabular}{c|ccc}
\hline	
k	&	$d=3$	&	$d=6$	&	$d=7 $	\\
\hline	
2	&	0.603	&	0.604	&	0.606	\\
3	&	0.584	&	0.517	&	0.502	\\
\hline	
\end{tabular}
\end{table}
In the Table~\ref{musu} we compare three cases $d=3, 6, 7$: $d=3$ -- the minimum, that is needed for parallelilsation with both methods, $d=7$ -- the case that was studied in previous section, $d=6$ -- to show the tendency for smaller $d$. We provide results obtained when the Rosenbrock function
of different dimensions $d=3, 6, 7$ was minimized by using our parallel modification of the 
Nelder-Mead method.  The values of the efficiency coefficients $\gamma_k$ are presented. 
They show that 
this parallel algorithm is quite stable and it is well-suited to be used in the three-level 
template solver for small dimension objective functions.

\begin{table}[ht]
\caption{The $\gamma_k$ values for the parallel Nelder-Mead algorithm from (Lee and Wiswall, 2007)
}
   \label{ju}
\centering
\begin{tabular}{c|ccc}
\hline	
k	&	$d=3$	&	$d=6$	&	$d=7 $	\\
\hline	
2	&	0.668	&	0.685	&	0.714	\\
3	&	--	    &	0.436	&	0.454	\\
4	&	--    	&	0.023	&	0.104	\\
5	&	--	    &	0.001	&	0.002	\\
6	&	--	    &	--	    &	0.001	\\
\hline	
\end{tabular}
\end{table}

Table~\ref{ju} presents results obtained by using the state-of-the-art parallel Nelder-Mead 
algorithm from \cite{lee2007parallel}.
It follows, that in all investigated cases the parallelisation degree is very limited, 
since the convergence drops significantly when the parallelization degree is 
increased. This method is mainly targeted to solve problems when the dimension of the 
objective function is big (e.g. problems in financial mathematics, when $d \approx 100$).


\section{Conclusions}\label{sec:conclusions}

In this paper we introduced
a three-level parallelisation template 
which utilises a new model-based load balancing which is based on experimental 
data. The technique was tested for three benchmarks. The experimental results 
confirmed the good accuracy of the new time prediction model. 

Comparing the three-level template  to the classical two-level scheme, the  
proposed scheme looks more promising  for development of efficient parallel 
algorithms in the case when a big number is computational resources is available. 

The possibilities of the three-level parallelization template are demonstrated 
for solving local optimization problems. 
On the first level a well-known Nelder-Mead 
algorithm was used. We proposed a family of parallel versions of this method, which
increases  the parallelisation degree up to the factor three. 
The proposed load balancing algorithm chooses the optimal version of the parallel
Nelder-Mead algorithm. It dynamically  
increases the parallelisation degree on the first level when the 
speed-up of the second and third levels begins to saturate.
 
For the considered test problem on the second level $M$ PDEs were solved numerically
and on the  third parallelisation level Wang's
algorithm was used to solve systems of linear equations. 
It was shown that there exists a limit 
for the speed-up that can be achieved due to limitations 
of Wang's algorithm. 
The proposed approach extends the parallelisation degree allowing to achieve 
an additional speed-up.

The proposed load balancing algorithm
limits the size of computational resources to preserve the efficiency 
requirement which can be controlled by selecting the parameter $E_{min}$.

\begin{acknowledgment}
Computations were performed using resources at the High Performance 
Computing Centre "HPC Sauletekis" in Vilnius University Faculty of Physics.
\end{acknowledgment}

\bibliography{amcs}


\makeinfo

\end{document}